# DESIGN AND PERFORMANCE ANALYSIS OF ULTRA LOW POWER 6T SRAM USING ADIABATIC TECHNIQUE

Mr. Sunil Jadav[1], Mr. Vikrant[2], Dr. Munish Vashisath[3]

[2]PG Student, Electrical & Electronics Engineering Dept. YMCAUS&T, Faridabad, Haryana

[1,3]Faculty, Electrical & Electronics Engineering Dept. YMCAUS&T, Faridabad, Haryana

`suniljadav1@gmail.com`[1], `vikrant_bmit@rediffmail.com`[2], `munish276@yahoo.com`[3]

***ABSTRACT:*** *Power consumption has become a critical concern in both high performance and portable applications. Methods for power reduction based on the application of adiabatic techniques to CMOS circuits have recently come under renewed investigation. In thermodynamics, an adiabatic energy transfer through a dissipative medium is one in which losses are made arbitrarily small by causing the transfer to occur sufficiently slowly. In this work adiabatic technique is used for reduction of average power dissipation. Simulation of 6T SRAM cell has been done for 180nm CMOS technology. It shows that average power dissipation is reduced up to 75% using adiabatic technique and also shows the effect on static noise margin.*

***Keywords****- Adiabatic Logic, Average Power dissipation, Static Noise Margin.*

## 1. INTRODUCTION

Adiabatic switching is a new approach for reducing power dissipation in digital logic. When adiabatic switching is used, the signal energies stored on circuit capacitances may be recycled instead of dissipated as heat [1]. For energy recovery circuit, the ideal energy dissipation when a capacitance C is charged from 0 to $V_{dd}$ or discharged from $V_{dd}$, through a circuit of resistance R during time T is given by

$$E = (RC/T)(V_{dd})^2$$

When T >> RC, the power consumption is much smaller than the conventional CMOS circuit, for which an energy of ½ C $(V_{dd})^2$ is required during a charge or discharge cycle.

If circuits can be rnade to operate in an adiabatic regime with consequently low energy dissipation, then the energy used to charge the capacitive signal nodes in a circuit may he recovered during discharge and stored for reuse[2]. The efficiency of such a circuit is then limited only by the 'adiabaticity' of the energy transfers. Conventional CMOS circuits are pathologically non adiabatic Capacitive signal nodes are rapidly charged and discharged (the energy transfer) through MOS devices (the dissipative medium). At times the full supply potential appears across the channel of the device, resulting in high device current and energy dissipation.

In this work a method based on adiabatic technique uses an ac power supply rather than dc for the recovery of energy [3]. Although adiabatic circuits consume zero power theoretically, they show energy loss due to nonzero resistance in the switches. The Simulation is carried out in 180nm CMOS technology using TANNER TOOLS.

## 2. 6T CMOS SRAM

As shown in Fig:1 the conventional 6T memory cell comprises of two CMOS inverters cross coupled with two pass transistors connected to a complementary bit lines. In Fig.1 the gate of access transistors NMOS3 and NMOS4 are connected to the wordline (WL) to have the data written to the memory cell from bit lines (BL). The bit lines act as I/0 buses which carry the data from memory cells to the sense amplifier. The main operations of the SRAM cells are the write, read and hold. The SNM is an important performance factor of hold and read operations, specifically in read operation when the wordline is '1' and the bit lines are precharged to '1'.

The internal node of SRAM which stores '0' will be pulled up through the access transistor and the drive transistor. This increase in voltage severely degrades the SNM during read operation. The read stability is mainly depends on the cell ratio. [4]

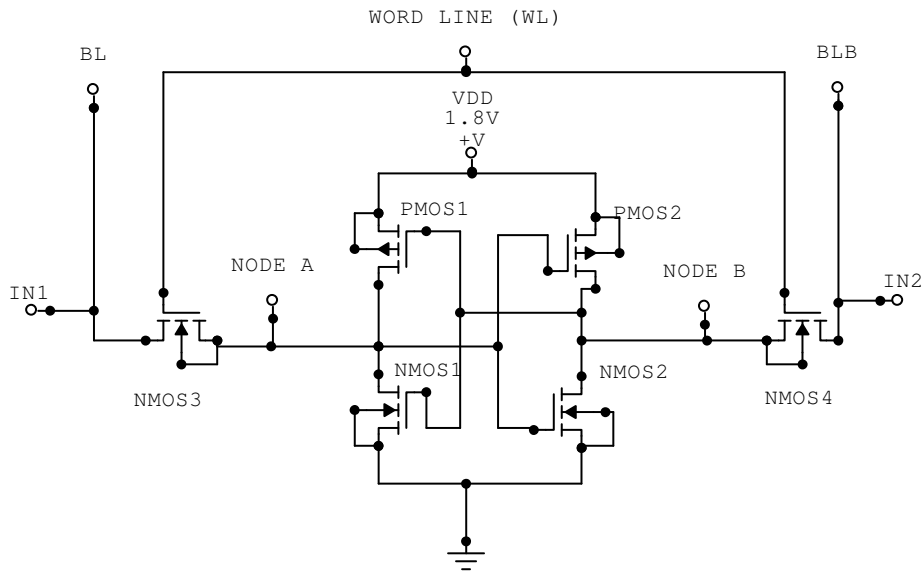

Figure 1 Conventional 6T CMOS SRAM Cell

### 2.1 Write Operation

For the write operation, in order to store logic '1' to the cell, BL is charged to Vdd and BLB is charged to ground and vise verse for storing logic '0'. Then the word line is switched to $V_{dd}$ to turn ON the NMOS access transistor. When the access transistors are turned ON, the values of the bitlines are written into Node A and Node B. The node which storing the logic '1' will not go to full Vdd because of voltage drops across the NMOS access transistor. After the write operation the wordline voltage is reset to ground to turn off the NMOS access transistor. The node with the logic '1' stored will be pulled up to full $V_{dd}$ through the PMOS driver transistors. Fig. 2 shown below shows the write '0' operation of 6T SRAM cell.

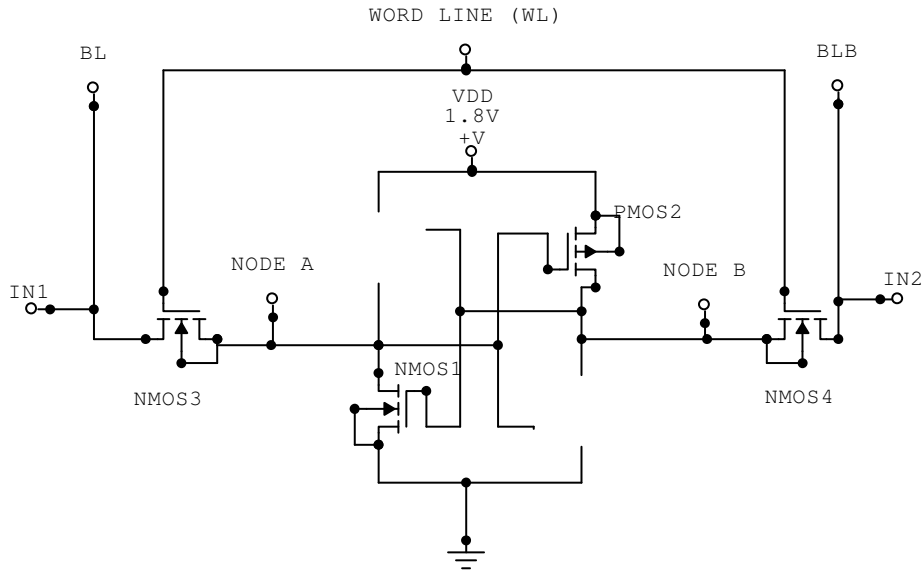

Figure 2 6T CMOS cell during write '0' operation.

## 2.2 Read Operation

For the read operation the bit lines and word lines are charged to $V_{dd}$. The node, storing logic '1' will pull the voltage on the corresponding bit lines up to a high (not $V_{dd}$ because of the voltage drop across the NMOS access transistor) voltage level. The sense amplifier will detect which bit line is at high voltage and which bit line is at ground.

## 2.3 Hold Operation

For hold operation the bitlines are charged to $V_{dd}$ and word lines connected to ground potential. The access transistors NMOS3 and NMOS4 disconnect the cell from the bit lines. The two cross-coupled inverters formed by PMOS1, PMOS2, NMOS1, NMOS2 will continue to reinforce each other as long as they are connected to the supply.

## 3. STATIC NOISE MARGIN OF SRAM

Noise margin can be defined using the input voltage to output voltage transfer characteristic (VTC). In general, Noise Margin (NM) is the maximum spurious signal that can be accepted by the device when used in a system while still maintaining the correct operation. If the consequences of the noise applied to a circuit node are not latched, such noise will not affect the correct operation of the system and can thus be deemed tolerable. It is assumed that noise is presented long enough for the circuit to react, i.e. the noise is static or dc. A SNM is implied if the noise is a dc source. An ideal inverter tolerates a change in the input voltage without any change in the output voltage until the input voltage reaches the switching point.

The static noise margin high and static noise margin low is defined as [5]

$$NM_H = V_{OH} - V_{IH}$$
$$NM_L = V_{IL} - V_{OL}$$

where $V_{IL}$ is the maximum input voltage level recognized as logical '0', $V_{IH}$ is the minimum input voltage level recognized as a logical '0', $V_{OL}$ is the maximum logical '0' output voltage, $V_{OH}$ is the minimum logical '1' output voltage.

### 3.1 Determination of SNM:

SNM is determined as a side of the maximum square drawn between the inverter characteristics [5]. An important advantage of this method is that it can be automated using a DC circuit simulator, which to a great degree extends its practical usefulness. In this approach an SRAM cell is presented as two equivalent inverters with the noise sources inserted between the corresponding inputs and outputs. Both series voltage noise sources ($V_N$) have the same value and act together to upset the state of the cell, i.e. they have an inverse polarity to the current state of each inverter of the cell. Applying the adverse noise sources polarity represents the worst-case equal noise margins. Fig.5 shows the superimposed normal inverter transfer curve of a read accessed 6T SRAM cell and its mirrored with respect to $x = y$ line counterpart in a $x-y$ coordinate system. This is a convenient arrangement. Since by knowing the diagonals of the maximum embedded squares we can calculate the sides.

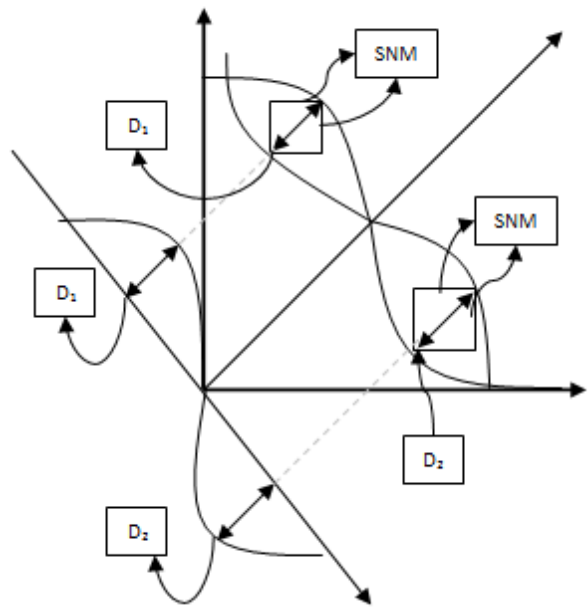

Figure 3 SNM estimation based on maximum Square.

## 4. ADIABATIC LOGIC TECHNIQUE

### 4.1 Conventional Charging

The dominant factor in the dissipation of a CMOS device is the *dynamic* power required to charge capacitive signal nodes within the circuit. This effect is illustrated in Fig. 4 for a simple CMOS inverter. To charge the signal node capacitance *C* from a supply of potential $V_{dd}$, a charge $q = CV_{dd}$ is taken from the supply through the P-type device. The total energy $E_T = QV_{dd} = C(V_{dd})^2$.

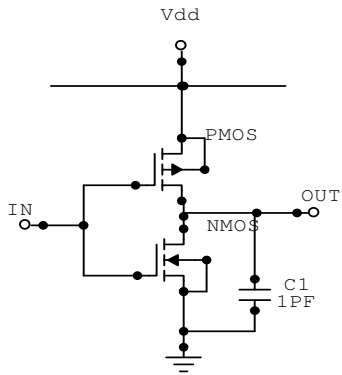

Figure 4 CMOS Inverter.

Only half of the energy is usefully applied to storing the signal on the capacitor-the other ½ C $(V_{dd})^2$ is dissipated as heat, primarily in the 'on' resistance of the p-type device. Note that the dissipation is independent of this resistance: it, is a result of the capacitor charge being obtained from a constant voltage source $V_{dd}$. The n-type device is used to discharge the ½ C $(V_{dd})^2$ energy stored in capacitor C by short circuiting the capacitor and dissipating energy as heat. Hence the total charge/discharge cycle has required an energy C $(V_{dd})^2$ - half being dissipated in charging and half being used for information storage before it too is dissipated during discharge.

### 4.2 Adiabatic Charging

Adiabatic switching can be achieved by ensuring that. The potential across the switching devices is kept arbitrarily small. The potential $V_r$ across the switch resistance is high in the conventional case because of the abrupt application of $V_{dd}$ to the RC circuit.

Adiabatic charging may be achieved by charging the capacitor from a time varying source that starts at $V_i = 0V$. The ramp increases towards $V_{dd}$ at a slow rate that ensures that $V_r = V_i - V_c$, is kept arbitrarily small. This rate is set by ensuring that its period $T \gg RC$.
In fact the energy dissipated is

$$E_{diss} = I^2RT = (CV_{dd}/T)^{2RT} = (RC/T) \, C \, (V_{dd})^2$$

A linear increase in *T* causes a linear decrease in power dissipation Adiabatic discharge can be arranged in a similar manner with a descending ramp.

Now if *T* is sufficiently larger than *RC,* energy dissipation during charging $E_{diss}$ → *0* and so the total energy removed from the supply is ½ $C(V_{dd})^2$ - the minimum required to charge the capacitor and hence hold the logic state. This energy may be removed from the capacitor and returned to the power supply during the discharge cycle if it too is performed adiabatically. As a result, given a suitable supply it should be possible then to charge and discharge signal node capacitances with only marginal net losses. Note that the *RC* time constant of a typical CMOS process is about ***100ps.*** If we set *T* to 10 time constants, the resulting delay through an adiabatic gate would be 1ns - making the gate viable in systems running with clock speeds in the tens to hundreds of megahertz range [2].

## 5. ADIABATIC 6T SRAM CELL

In this proposed SRAM cell adiabatic technique is used. In this adiabatic technique ac power supply is used. By using the ac power supply rather than dc the average power dissipation is reduced. The Fig 5 shown below shows the adiabatic 6T SRAM cell. Adiabatic switching can be achieved by ensuring that the potential across the switching devices is kept arbitrarily small. the potential $V_r$ across the switch resistance is high in the conventional case because of the abrupt application of $V_{dd}$ to the RC circuit.

Adiabatic charging may be achieved by charging the capacitor from a time varying source that starts at $V_I$= 0V to $V_{dd.}$ For this purpose an AC power supply is used.

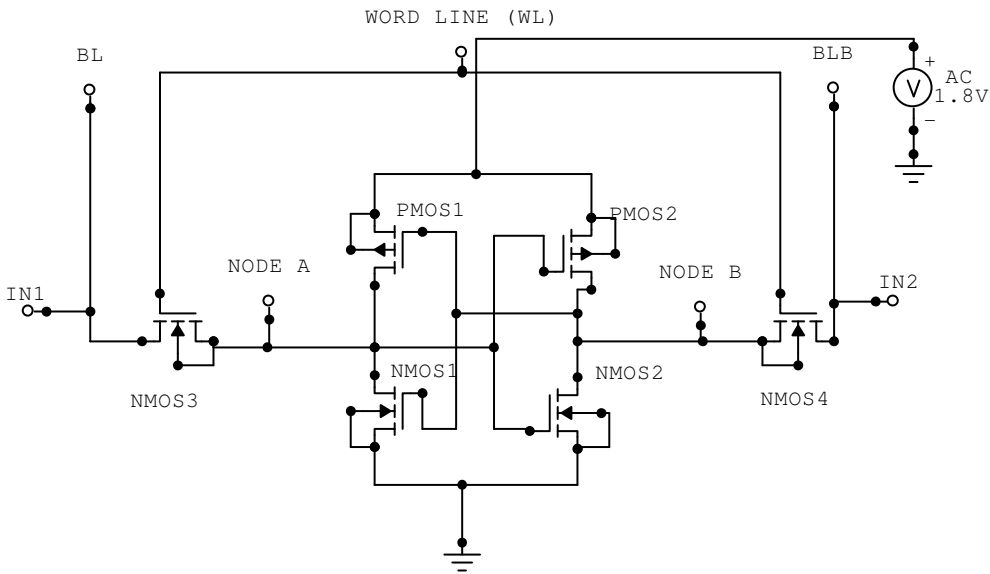

Figure 5 6T Proposed SRAM cell

## 6. SIMULATION RESULTS

In this section comparison of conventional 6T SRAM and adiabatic 6T SRAM Cell for different operation of SRAM Cell on the basis of average power dissipation and static noise margin has been carried out. In general DC power supply is used. In adiabatic logic AC power supply is used rather than DC power supply [2].

Tables and figure shown below shows the comparison of average power dissipation for conventional and adiabatic SRAM with different operations.

Table: I show the comparison of average power dissipation during write '0'/ '1' between conventional and adiabatic SRAM cell. The average power dissipation is reduced up to 87% using adiabatic logic technique during write operation.

Table I

| SRAM | Average Power Dissipation During Write '0'/'1' |
|---|---|
| Conventional | 1.59E-05 |
| Adiabatic | 2.07E-06 |
| % Decrease | 87% |

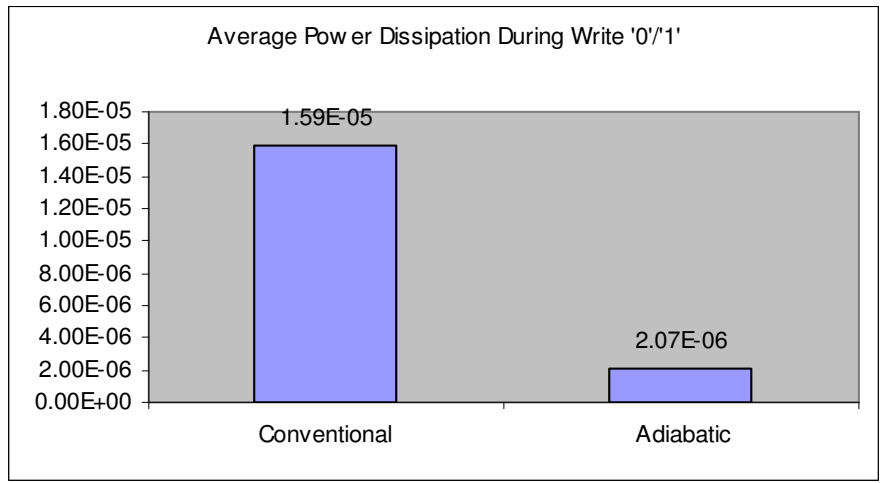

Figure 6 Comparison graph of Average Power Dissipation During Write '0'/ '1'

Figure 6 shows the graphical representation of average power dissipation during write '0' and write '1' operation. The average power dissipation during these operation is reduced up to 87%.

Table II

| SRAM | Average Power Dissipation During Write/Hold |
|---|---|
| **Conventional** | 7.93E-06 |
| **Adiabatic** | 2.64E-06 |
| **% Decrease** | 66.75 |

Table II shows the comparison of average power dissipation during Write/Hold operation between conventional and adiabatic SRAM cell. The average power dissipation is reduced up to 67% using adiabatic logic technique during write/hold operation.

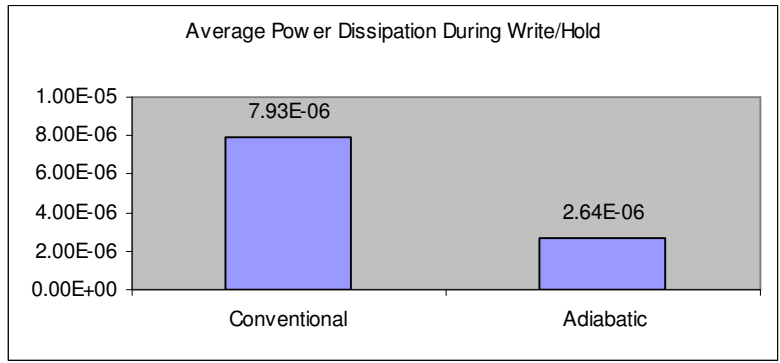

Figure 7 Comparison graph of Average Power Dissipation During Write/Hold

Figure 7 shows the graphical representation of average power dissipation during write and hold operation of SRAM.

Table: III

| SRAM | Average Power Dissipation During Write/Read |
|---|---|
| **Conventional** | 1.46E-05 |
| **Adiabatic** | 2.22E-06 |
| **% Decrease** | 84.78 |

Table III shows the comparison of average power dissipation during Write/Read operation between conventional and adiabatic SRAM cell. The average power dissipation is reduced up to 85% using adiabatic logic technique during write/read operation.

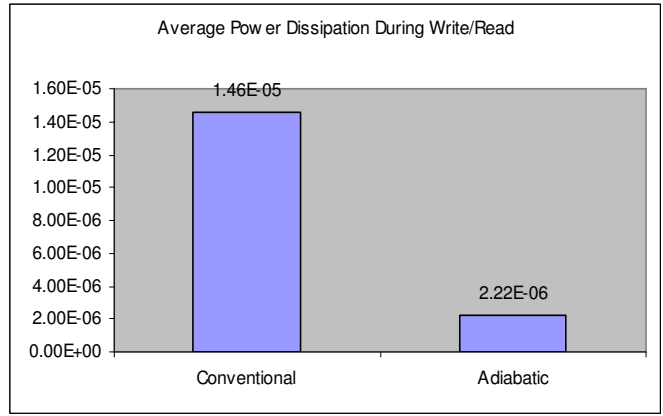

Figure 8 Comparison graph of Average Power Dissipation During Write/Read

Figure 8 shows the graphical representation of average power dissipation during write and read operation.

Table: IV

| Transistor | During Write/Hold Operation | | |
|---|---|---|---|
| | Conventional SRAM | Adiabatic SRAM | % Reduced |
| NMOS 3 | 4.97E-10 | 3.25E-10 | 34.61 |
| NMOS 4 | 6.40E-10 | 4.97E-10 | 22.34 |

Table IV and Fig.9 shows the leakage current component of conventional and adiabatic SRAM. It concludes that the leakage current is reduced using adiabatic logic technique.

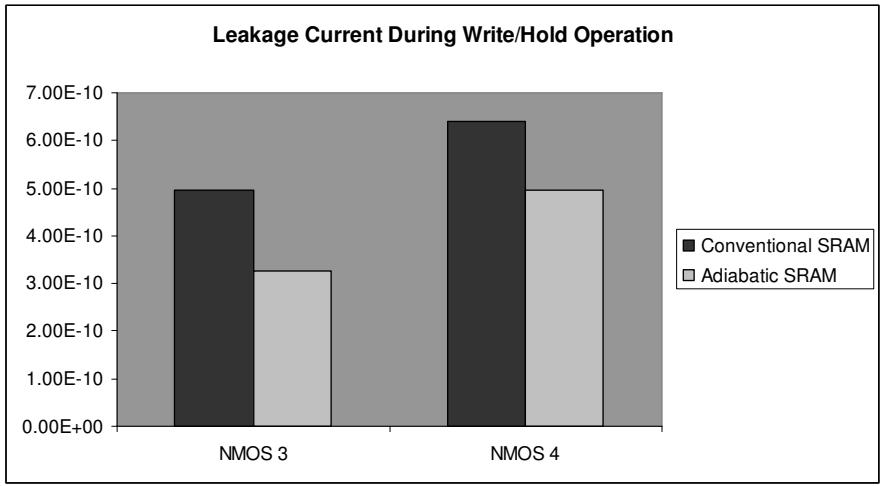

Figure 9 Comparison Graph of Leakage Current with and without Low Power Technique during Write/Hold Operation

International Journal of VLSI design & Communication Systems (VLSICS) Vol.3, No.3, June 2012

Figure shown below shows the butterfly curve for calculation of Static Noise Margin. Table: IV shown below shows the comparison of static noise margin of conventional and adiabatic 6T SRAM cell.

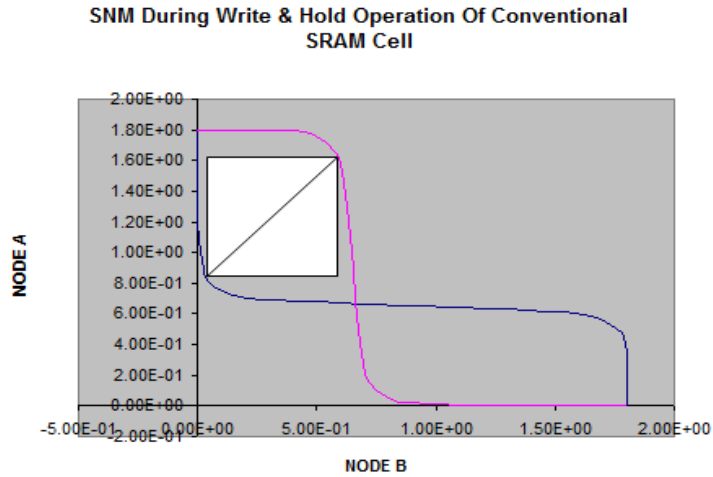

Figure 10 Butterfly Curve for Static Noise Margin of Conventional 6T SRAM Cell

Figure 10 shows the butterfly curve of conventional SRAM for calculating the static noise margin during write and hold operation.

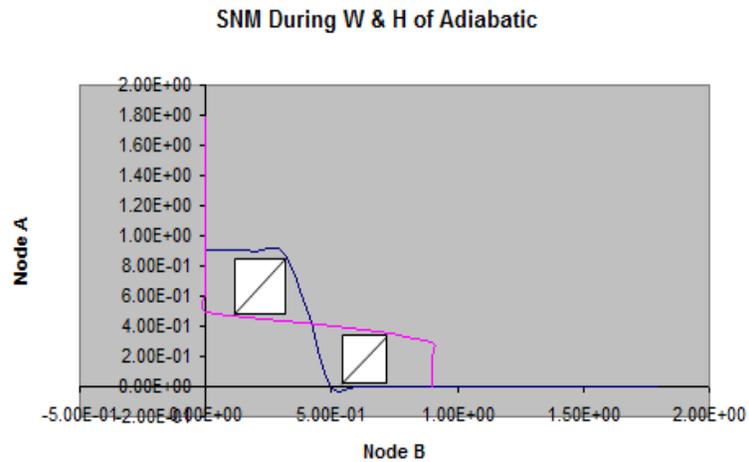

Figure 11 Butterfly Curve for Static Noise Margin of Adiabatic 6T SRAM Cell

Figure 11 shows the butterfly curve of adiabatic SRAM for calculating the static noise margin during write and hold operation. SNM is reduced using adiabatic logic technique.

Table V

| SRAM | Static Noise Margin |
| --- | --- |
| Conventional | 1.13E+00 |
| Adiabatic | 5.66E-01 |

104

Table V shows that the value of Static noise margin of adiabatic SRAM is reduces as compare to conventional 6T SRAM.

## 7. CONCLUSION

In this work adiabatic technique is used for reduction of average power dissipation with no performance degradation. Simulation of 6T SRAM cell has been done for 180nm CMOS technology. By using this technique the average power consumed is reduced up to 87% during write operation, during write and hold operation power is reduced up to 66% and during write and read operation average power consumed is reduced up to 85%. The static noise margin is also reduced by using adiabatic technique. By using Adiabatic technique for design of SRAM cell the average power dissipation is reduced with no performance degradation. In future, techniques to improve static noise margin for adiabatic logic technique would be carried out.